# Deadline-Driven Multi-node Mobile Charging

Xunpeng Rao*, Panlong Yang*, Haipeng Dai†, Tao Wu‡, Hao Zhou*, Jing Zhao§, Linlin Chen§, and Peng-Jun Wan§,
*University of Science and Technology of China
†Nanjing University
‡Army Engineering University of PLA
§Department of Computer Science, Illinois Institute of Technology, Chicago, IL 60616 USA
Email: raoxunpeng@gmail.com, plyang@ustc.edu.cn, haipengdai@nju.edu.cn, kitewind@ustc.edu.cn,
terence.taowu@gmail.com, jzhao29@hawk.iit.edu, lchen96@hawk.iit.edu, wan@cs.iit.edu

*Abstract*—Due to the merit without requiring charging cable, wireless power transfer technologies have drawn rising attention as a new method to replenish energy to Wireless Rechargeable Sensor Networks (WRSNs). In this paper, we study mobile charger scheduling problem for multi-node recharging with deadline-series. Our target is to maximize the overall effective charging utility, and minimize the traveling time as well. Instead of charging only once over a scheduling cycle, we incorporate the multiple charging strategy for multi-node charging with deadline constraint, where charging spots and tour are jointly optimized. Specifically, we formulate the effective charging utility maximization problem as to maximize a monotone submodular function subject to a partition matroid constraint, and propose a simple but effective $\frac{1}{2}$-approximation greedy algorithm. After that, we present the grid-based *skip-substitute* operation further to save the traveling time, which can increase the charging utility. Finally, we conduct the evaluation for the performance of our scheduling scheme. Comparing to the *Early Deadline First* scheme, the simulation and field experiment results show that our algorithm outperform EDF by 37.5% and 37.9%, respectively.

## I. INTRODUCTION

### A. Background and Motivation

Since Kurs *et al.* [1] show that the energy can be transmitted wirelessly, Wireless Power Transfer (WPT) has drawn increasing attention in industry and academic circles for the merit of requiring no charging cable. In order to promote the standardization of WPT, many organizations have jointed the Wireless Power Consortium [2], such as Qualcomm, Samsung, Huawei, *etc*. Many applications have been promoted including structures monitoring [3] and body sensor network [4].

The battery powered Wireless Rechargeable Sensor Networks (WRSNs) have been benefited a lot from WPT. Many mobile charger scheduling solutions have been proposed for sensing and communication tasks [5]–[9]. In addition, mobile charging could improve the power supply reliability in dealing with instability of harvesting ambient energy [10]. Despite tremendous amount of academic researches have been proposed for mobile charging network, they are still struggling to come to reality. For example, time constraints should be respected for time critical tasks [11]. Most of the solutions assumed that the mobile charger can always meet the time requirement for charging. While in reality, the energy requirements are not always met due to time budget [5], effects of charging distance, angle [6] [12], *etc*.

Most of existing researches [5] [8] study the mobile charger scheduling with single-node mode, and then formulate the problem into the typical traveling salesman problem or orienteering problem. However, such formulations under-utilize the ability of WPT (*e.g.*, charging wirelessly, multi-node charging), and cannot apply to the case of multi-node charging mode. In multi-node case, the Xie *et al.* [9] have shown that the advantage of multi-node charging in a dense wireless sensor network. However, it is hardly to apply to time critical sensor networks. In multi-node case, the charging spots selection with deadline constraint is a serious issue, especially the dense networks. Furthermore, the typical problem formulations would result in that all nodes own the charging chance only once over a scheduling cycle.

In summary, there are two deficiencies in existing scheduling with time restrictions. First, the single-node charging scheme with deadline constraint cannot apply to the multi-node charging case. Especially in the case of node-intensive deployment, it is hardly to ensure that each node would be charged in a timely manner. Second, the existing problem formulations result in an imperfect charging scheme, that is all nodes are allowed to charge only once (one chance charging) over a scheduling cycle. However, in real scenario, the charging scheduling should be very flexible to maintain the effective charging tasks. In this paper, we study the <u>M</u>ulti-node <u>MO</u>bile Cha<u>R</u>ging Towards D<u>E</u>adline-series (MORE) problem. In that, we compensate the deficiencies for existing problem formulation by extending the one by one charging scheme to multi-node charging scheme in deadline-driven scheduling. In addition, we remove the imperfect and acquiescent one chance charging scheme by allowing the nodes be charged more than once over a scheduling cycle.

### B. Challenges and Contributions

It is non-trivial when time delay is respected and fully evaluated during mobile charger scheduling. Particularly, there are two fundamental challenges need to be formally addressed before achieving a deadline-driven scheduling.

First, time constraints are usually independent to the spatial constraints. If a charger could not reach a node before deadline, it won't get any rewards. Thus, it is difficult for energy-saving concern to apply the conventional method such as "best effort" schemes to our scenario.

Second, the deadline aware charger scheduling needs to jointly optimize the charging spots and tour. Unfortunately, even solving either single problem is NP-hard. Moreover, decomposing these two steps would lead to possible performance loss, which further complicates our problem.

In tackling aforementioned challenges, we consider the relaxed MORE by spatial and temporal discretization. Thus, we could formulate the problem into a monotone submodular function maximization subject to a partition matroid constraint, and propose a simple but effective $\frac{1}{2}$-approximation greedy algorithm. After that, we present the *skip-substitute* operation to save the traveling time, which can increase the charging utility. In summary, our contributions could be summarized as follows:

- To the best of our knowledge, we are the first to incorporate multiple charging strategy for multi-node charging with deadline constraint, where charging spots and tour are jointly optimized.
- We decompose the complicated problem into two steps. First, after the space and temporal domain decomposition, a near optimal scheme with approximation ratio of 1/2 for MORE-R is available for charging locations selection. Second, the charging tour is further optimized to save traveling time by the proposed *skip-substitute*.
- We conduct extensive evaluations and real deployment to validate the proposed schemes, which outperforms the classic "*Early Deadline First (EDF)*" scheme [13] by 37.5% and 37.9% under simulation and real testbed evaluation, respectively.

*C. Paper organization*

The remainder of this paper proceeds as follows. We first review the related work about mobile charging scheduling in Sec. II. Then we present the problem MORE in Sec. III. Next, we propose the maximizing charging utility scheme in Sec. IV, and constructing traveling route scheme in Sec. V. In Sec. VI and Sec. VII, we conduct extensive simulations and field experiments, respectively. Finally, we conclude our work in Sec. VIII.

## II. RELATED WORK

Since the wireless power transfer has drawn increasing attentions, the researches [5]–[9], [14]–[21] on scheduling mobile chargers to serve WRSNs have been for several years. Existing studies can be divided into two broad types: single-node charging and multi-node charging.

**Single-node charging:** Shi *et al.* [14] gave an optimal charging scheduling of maximizing the proportion of vacation time over a cycle to periodically serve nodes one by one. Moreover, Rao [6] studied the periodic scheduling scheme by considering the effects of charging angle and distance on charging efficiency simultaneously. Given heterogenous charging frequency, Xu *et al.* [19] investigated how to schedule multiple chargers for multiple charging cycle to minimize the traveling distance and charging delay. Chen *et al.* [5] propose solution to maximize the number of charged nodes with time budget. The similar work arises in Ye's work [8], where the charger is scheduled to maximize the charging utility under the time window restricts. Conclusively, these aforementioned solutions cannot be applied to the mobile charging problem with bounded charging disk model for the charging points selection problem should be great respected.

**Multi-node charging:** Tong *et al.* [21] studied the effects of multi-node wireless charging on deployment for sensor nodes to maintain the data transmission. The periodic charging scheme in Shi previous work was extended to the case of multi-node charging with the bounded charging range [9]. Xie *et al.* [18] studied the problem of bundling WPT and mobile sink, where the stopping points, charging schedule, data flow routing are jointly optimized, to minimize the energy consumption over the entire system. Furthermore, the fine-grained charging points selection by area partition were studied in Xie later work [16]. The case of multi-node mobile charging with unbounded charging model was studied by Fu *et al.* [7], which aimed at searching charging spots to minimize the charging delay by ignoring the traveling time. Conclusively, there was little work in deadline-driven multi-node mobile charging. The charging points selection scheme would be affect a lot for the mismatch between deadline constraints and spatial constraints. Therefore, these aforementioned solutions are hardly to directly handle the case with deadline constraints.

In summary, most of existing studies focus on scheduling charger to meet different optimal objectives without considering the delay sensitive applications. Unlike these mentioned studies, we here study the mobile charger scheduling of deadline-driven multi-node mobile charging. Furthermore, we propose an efficient algorithms to minimize the traveling time.

## III. PROBLEM STATEMENT

*A. Network Model and Charging Model*

Suppose we have $N$ stationary rechargeable sensor nodes densely distributed on a two-dimension plane $\Omega$. We denote $n$ as a typical $n^{th}$ node, and $p_n$ as the location in $\Omega$. A mobile charger equipped with omnidirectional antennas is employed to wander with constant speed $V$ in the plane to charge the nodes wirelessly. The charger starts from a depot $S$, and it should return to the depot after finishing the charging tasks. There are $K$ candidate stop locations for charger in plane $\Omega$ ($K$ is a fairly large number). Similarly, we denote $k$ as the $k^{th}$ stop location for charger, and $p_k$ as the location in $\Omega$.

In this work, we use the charging model proposed in [15] as follows:
$$P(d) = \begin{cases} \frac{\alpha}{(d+\beta)^2}; & d \leq D \\ 0; & d > D \end{cases}$$

where $d$ is the charging distance between the charger and a node, $\alpha$, $\beta$ are known constant environment parameters depending on the hardware design of charger, and $D$ is the maximum charging distance. Under the above charging model, we know that the nodes can only be charged within the effective charging range, and the received power at nodes is

TABLE I: Annotations for frequently used symbols

| Symbol | Definition |
|---|---|
| $N$ | Number of rechargeable sensor nodes |
| $D$ | The threshold of charging distance |
| $P_{kn}$ | The charging power from position $p_k$ to $p_n$ |
| $\tau_n$ | The charging deadline of node $n$ |
| $e_n$ | The charging demands of node $n$ |
| $T$ | The time budget of scheduling |
| $X_k^n$ | The variable to describe the state of charging |
| $\delta$ | The length of grids side |
| $\lambda$ | The approximation error after grid partition |
| $\Delta t$ | The length of each time slot |
| $T_s$ | The set of time slots $T_s = \{s_1, s_2, ..., s_m\}$ |
| $G$ | The set of grids $G = \{g_1, g_2, ..., g_\Gamma\}$ |
| $Q_n$ | The effective charging energy at node $n$ |

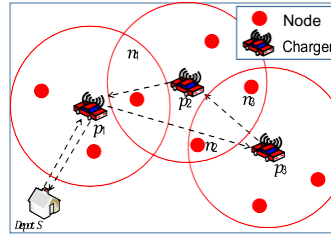
Fig. 1: Mobile charging scenario.

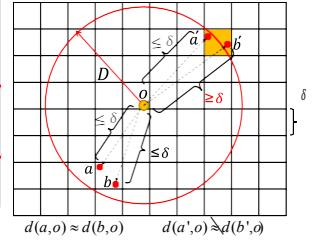
Fig. 2: Grid partition.

nearly inversely proportional to the squared charging distance. For simplification, we use $d_{k,n}$ to denote the distance $d(p_k, p_n)$ between $p_k$ and $p_n$.

Moreover, we assume that node $n$ issues the charging demand of $\phi_n = (\tau_n, e_n)$, where $e_n$ denotes the amount of required energy for node $n$, and $\tau_n$ denotes the corresponding charging deadline. Without loss of generality, we assume that $Y = \{\tau_1, \tau_2, ..., \tau_N\}$ is sorted in ascending sequence (i.e., $\tau_i \leq \tau_j, i \neq j$). Let $T$ denote the maximum deadline in Y, i.e., $T = \tau_N$. In this work, we assume that the charging requirement including the required charging energy and deadlines are known to the charger in priori.

### B. Charging Utility Model

Based on the aforementioned model, we consider the effective charging energy to describe the valid charging energy received at nodes. The effective charging energy for $n$ is the amount of charging energy received at nodes before the deadline $\tau_n$. Note that, even the nodes may be charged after deadline, we consider such extra amount of energy as invalid. Therefore, we propose the following linear bounded *charging utility* model to capture such bounded charging demand,

$$U_n(Q_n) = \begin{cases} \frac{1}{e_n} \times Q_n, & Q_n \leq e_n; \\ 1, & Q_n > e_n, \end{cases}$$

where $Q_n$ denotes the received charging energy before deadline, and $e_n$ is the charging demand of node $n$. In that, the charging utility is proportional to the received charging energy when $Q_n$ less than required demand, and then charging utility is a constant when $Q_n$ exceeds the threshold $e_n$.

Next, we give the computation of charging energy $Q_n$ for node $n$ over the deadline budget $T$ as

$$Q_n = \sum_{k \in \Omega} P(d_{k,n}) \times \min\{\tau_n - r_k, t_k\} \times X_k^n,$$

where $t_k$ denotes the duration for charger at $k$, $r_k$ is the arriving time at $k$ ($r_k = \sum_{i=1}^{k-1} t_i + \sum_{i=1}^{k-1} d_{i,i+1}/V$), and $X_k^n$ is the indicator function to describe whether the scheduling time exceeds the deadline $\tau_n$ at the $k$ or not, which is

$$X_k^n = \begin{cases} 1, & r_k < \tau_n; \\ 0, & otherwise. \end{cases}$$

### C. Problem Formulation

Due to the charging deadline, it is intuitive to schedule the mobile charger to maximize the charging utility under the constrained deadline. Based on the proposed model in last subsection, the Multi-node MObile ChaRging Towards DEadline-series (MORE) problem can be formulated as:

$$\max_{\{t_k, p_k\}} \sum_{n=1}^{N} U_n(Q_n)$$
$$s.t. \sum_{k \in \Omega} t_k \leq T.$$

An feasible path in Fig. 1, the charger is scheduled to stop at $p_1, p_2, p_3$ to serve multi-nodes simultaneously. In that, the multiple charging scheme is illustrated well, that the node $n_1$ is charged when charger at $p_1$ and $p_2$ which means that $n_1$ is charged for two times. Similarly, the node $n_2$ and $n_3$ are charged for two times.

### D. Problem Hardness Analysis

In the optimization problem MORE, we need to figure out the duration $t_k$ for each stop locations $k \in \Omega$. Otherwise, the boolean variable $X_k^n$ is determined by the previous durations and traveling time. In order to maintain the sensing or communication tasks driven by energy with best efforts, the deadline aware charging scheduling needs the jointly optimization for the charging spots and tour. Unfortunately, even solving either single problem is NP-hard. Moreover, decomposing these two steps would lead to possible performance loss, which further complicates our problem. The most straightforward way is to compute all stop locations and corresponding durations. However, this enumeration method would incur very high computational complexity. Therefore, we first consider a relaxed version of MORE, which aims at finding the stop locations for charger to maximize the charging utility in Sec. IV. To handle this relaxed version, we introduce the area discretization and time partition in slot methods to give candidate stop locations and charging time with guaranteed approximation ratio. After that, we propose a *skip-substitute* technique to further optimize the grid-based traveling tour in Sec. V.

## IV. SOLUTION TO MORE-R

In this section, we consider a relaxed version of MORE (MORE-R for short), that is, finding the optimal stop locations for the charger to maximize the charging utility by ignoring the traveling time. Similar to [7] [16], we assume that, MORE-R itself is meaningful for applications such as dense sensor networks where the charging time is predominant and traveling time is negligible. First, we use an area discretization technique to get the candidate stop locations with guaranteed approximation ratio. Further, we use an time discretization technique to transform MORE-R into a monotone submodular function maximization subject to partition matroid constraint. After that, we present an efficient greedy algorithm with approximation ratio of 1/2.

*A. Area Discretization*

We consider $\Omega$ as a square plane. As any point in $\Omega$ can be chosen as a candidate location, which implies that the number of candidate locations is infinite, we present an area discretization technique to reduce the solution space from infinite to finite. As shown in Figure 2, we discretize the plane $\Omega$ by grid partition, in which the side length of each grid is set to be $\delta$. Suppose we obtain $\Gamma$ number of grids which are dentoed by $g_i(i = 1, 2..., \Gamma)$, where $\Gamma = \frac{|\Omega|}{\delta^2}$. Then the plane $\Omega$ can be clustered into the number of grids. Next, we approximately regard that all locations in a grid are identical for candidate stop location selection. That is, we randomly choose a point in each grid as the candidate stop location for charger. For the example in Figure 2, a device is deployed at location $o$ and its charging range is $D$. For the grid in which point $b$ locates, we take $b$ as the candidate stop location for this grid. Thus even a charger stops at another point $a$, we approximately regard it is at point $b$. Consequently, the distance $d(a, o)$ between point $a$ and $o$ can be approximated as $d(b, o)$, i.e., $d(a, o) \approx d(b, o)$.

The above approximation will result in error of charging power received at nodes. To bound such error, one critical issue is to ensure that the original distance and the approximated distance should be either both within threshold $D$ or both not because otherwise, the non-zero charging power may be approximated as zero charging power or opposite. For example, in Figure 2, the gird in orange is divided into two parts by the charging threshold circle. Obviously, the distance $d(a^l, o)$ between point $a^l$ and $o$ cannot be approximated by $d(b^l, o)$ for the former leads to nonzero power while the latter doesn't. Therefore, we modify the charging model by replacing distance between a node and a charger with distance between a node and a grid, to guarantee that the charging model is not violated with the approximated charging distance. Therefore, the charging model can be expressed as

$$P(d(g(a), o)) = \begin{cases} \frac{\alpha}{(d(g(a),o)+\beta)^2}, & d(g(a), o) \leq D; \\ 0, & \text{otherwise}, \end{cases}$$

where $d(g(a), o)$ is the distance between the grid $g(a)$ including $a$ between node $o$, which depends on the longest distance from $o$ to $g(a)$. We consider the charging power as a constant for node $o$ when charger at grid $g(a)$. That is to say, $P(d(a, o))$ is approximated as $P(d(g(a), o))$. For this kind of approximation, we have the following Lemma.

**Lemma 1.** *Approximated charging power: Given $\delta = \frac{\sqrt{2}}{2}\beta(\sqrt{\frac{1}{1-\lambda}}-1)$, where $0 \leq \lambda \leq 1$, we have the approximation ratio of charging power as*

$$P(d(g(a), o)) \geq (1-\lambda)P(d(a, o)).$$

*Proof:* Let $d$ denote the original distance between $a$ and $o$, and $d^l$ denote the approximated distance between $g(a)$ and $o$. Then we have

$$P = \frac{\alpha}{(d+\beta)^2}, \quad P^l = \frac{\alpha}{(d^l+\beta)^2}$$

Given $\delta = \frac{\sqrt{2}}{2}\beta(\sqrt{\frac{1}{1-\lambda}}-1)$ and $0 \leq \lambda \leq 1$, we have

$$\frac{P^l}{P} = \left(\frac{d+\beta}{d^l+\beta}\right)^2 \geq \left(\frac{d+\beta}{d+\beta+\frac{\sqrt{2}}{2}\cdot 2\delta}\right)^2 = \left(\frac{d+\beta}{d+\frac{\sqrt{\beta}}{\sqrt{1-\lambda}}}\right)^2$$

$$= (1-\lambda)\left(\frac{d+\beta}{d\sqrt{1-\lambda}+\beta}\right) \geq (1-\lambda)$$

Thus, we get $P^l \geq (1-\lambda) \cdot P$. ∎

After the area discretization, we can obtain the candidate stop locations set $G = \{g_1, g_2, ..., g_\Gamma\}$.

*B. Time Domain Discretization*

In this subsection, we present a time domain discretization technique for analysis. In particular, we divide the time domain into uniform slots with duration of $\Delta_t$, and require that the charger should stop at one single location during a certain time slot. Clearly, the total number of unit durations is given by $m = \frac{T}{\Delta_t}$, and we denote the set of time slots as $T_s = \{s_1, s_2, ..., s_m\}$. Next, for node $n$, its deadline $\tau_n^l$ is approximated as $\tau_n^l = \lceil\frac{\tau_n}{\Delta_t}\rceil \cdot \Delta_t$, which can be bounded with $|\tau_n - \tau_n^l| < \Delta_t$. After this discretization, each node can be charged with duration of no less than $\Delta_t$, and the deadlines of nodes will never appear inside a time slot.

Let $\chi$ denote the set of stop grids for the charger before deadline budget $T$, and then we can get the effective charging energy at node $n$

$$Q_n(\chi) = \sum_{g_k \in \chi} P(d(g_k, p_n)) \times \Delta_t \times X_k^n,$$

where $g_k$ is the stop grid in $\chi$ at the $k^{th}$ slot $s_k$ for charger. Accordingly, the indicator function $X_k^n$ can be rewritten as

$$X_k^n = \begin{cases} 1, & (k-1)\cdot\Delta_t < \tau_n^l; \\ 0, & (k-1)\cdot\Delta_t \geq \tau_n^l. \end{cases}$$

*C. Charger scheduling graph*

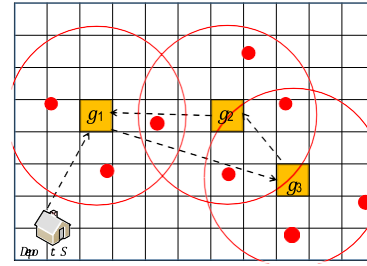
Fig. 3: The scenario after discretization.

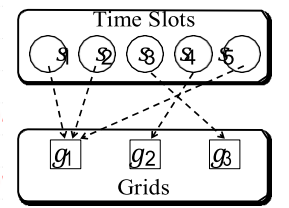
Fig. 4: Charger scheduling graph.

We define a bipartite graph $M$ as charger scheduling graph, with vertex set

$$T_s := \{s_1, s_2, ..., s_m\},$$

$$G := \{g_1, g_2, ..., g_\Gamma\},$$

edge set
$$E := \{a_{11}, a_{12}, ..., a_{1\Gamma}, ..., a_{m1}, a_{m2}, ..., a_{m\Gamma}\}.$$

We claim that a subset of edges $E$, denoted as $\chi$, is valid if there is no common vertex in $T_s$ for any two edges in $\chi$. Any valid $\chi$ can be mapped to a charger scheduling solution for the MORE-R problem if edge $a_{ij}$ represents the event that the charger stops at grid $g_j$ in time slot $s_i$.

We give an example by discretizing the scenario in Fig. 1. Fig. 3 and 4 show the charger scheduling graph $A$ consists of three grids $g_1$, $g_2$, $g_3$ and five slots $s_1$, $s_2$, $s_3$, $s_4$, $s_5$. In that, the feasible charger scheduling path is $S \to g_1 \to g_3 \to g_2 \to g_1 \to S$ denoted by dotted line, where the edge set of $\chi$ is $\{a_{11}, a_{21}, a_{33}, a_{42}, a_{51}\}$.

Now, the MORE-R problem is converted to choosing a valid $\chi \subseteq E$ to maximize the object function as $\sum_{n=1}^{N} U_n(Q_n(\chi))$.

### D. Theorems about subgraph $\chi$ and $\sum_{n=1}^{N} U_n(Q_n(\chi))$

**Theorem 1.** $(E, \chi)$ *is a partition matroid for valid $\chi$.*

*Proof:* In a bipartite graph $M = (T_s, G, E)$, one may form a matroid where the elements are the vertices on the side of bipartite.

Next, we partition the edge set $E$ into $m$ disjoint subsets, $E = \cup_{i=1}^{m} E_i^l$, where the subset $E_i^l = \{a_{i1}, a_{i2}, ..., a_{i\Gamma}\}$. Note that, $\chi$ is a feasible scheduling, if and only if $|\chi \cap E_i^l| \leq 1$, for charger can only stop at one grid in each slot. Thereby, we give the family of independent feasible sets $\chi$ by

$$\chi = \{I : I \subseteq E, |I \cap E_i^l| \leq 1, i = 1, 2, ..., m\}.$$

Thus, $M = (E, \chi)$ is a partition matroid. ∎

**Theorem 2.** $\sum_{n=1}^{N} U_n(Q_n(\chi))$ *is monotone submodular set function for $\chi$.*

*Proof:* Firstly, note that the set function $\sum_{n=1}^{N} U_n(Q_n(\chi))$ is the sum of a number of $U_n(Q_n(\chi))$. A favourable property of submodularity in [22] shows that the function consists of sum of a number of independent submodular functions is submodular. Due to the favourable additivity of submodularity, we only need to prove the submodularity of $U_n(Q_n(\chi))$. Before the proof for submodularity of $U_n(Q_n(\chi))$, we give an useful lemma.

**Lemma 2.** *Given a set submodular function $f(\cdot)$ defined on limited set $V$ and a monotone incremental concave function $t(x)$ which meets $t(0) = 0$, the compound function $g(\cdot) = t(f(\cdot))$ is submodular.*

It is intuitive that the charging utility model is a concave function. Therefore, with the Lemma 2, we only need to prove that the set function $Q_n(\chi)$ is submodular. Next, we are going to prove that the set function $Q_n(\chi)$ is monotonously nondecreasing submodular.

First of all, it is obvious that $Q_n(\varphi) = 0$. Second, we consider two arbitrary subsets $A$ and $B$, subject to $A \subseteq B \subseteq \chi$.

We use $P_{kn}$ to denote $P(d(g_k, p_n))$, then we have,
$$Q_n(B) - Q_n(A) = (\sum_{g_k \in B} X_k^n \cdot P_{kn} - \sum_{g_k \in A} X_k^n \cdot P_{kn}) \cdot \Delta_t$$
$$= \sum_{g_k \in B \setminus A} X_k^n \cdot P_{kn} \cdot \Delta_t \geq 0.$$

Therefore, $Q_n(\chi)$ is monotonously nondecreasing.

Finally, consider an arbitrary element $u \in \chi \setminus B$. Let $\Delta_Q(u|A) = Q_n(A \cup \{u\}) - Q_n(A)$. When adding the $u$ into the subsets $A$ and $B$, the increment are
$$\Delta_Q(u|A) = (\sum_{g_k \in A \cup \{u\}} X_k^n \cdot P_{kn} - \sum_{g_k \in A} X_k^n \cdot P_{kn}) \cdot \Delta_t$$
$$= X_{|A|+1}^n \cdot P_{|A|+1,n} \cdot \Delta_t;$$
$$\Delta_Q(u|B) = (\sum_{g_{k'} \in B \cup \{u\}} X_{k'}^n \cdot P_{k'n} - \sum_{g_{k'} \in B} X_{k'}^n \cdot P_{k'n}) \cdot \Delta_t$$
$$= X_{|B|+1}^n \cdot P_{|B|+1,n} \cdot \Delta_t.$$

Note that $|A| \leq |B|$, then $|A| + 1 \leq |B| + 1$. Furthermore, it is obvious that $X_k^n \geq X_{k'}^n$ if and only if $k' \geq k$. Thus, we get $X_{|B|+1}^n \leq X_{|A|+1}^n$ and
$$\Delta_Q(u|B) \leq \Delta_Q(u|A).$$

In summary, the set function $Q_i(\chi)$ is monotonously nondecreasing submodular. ∎

### E. Problem Transformation

Thereby, our optimization problem can be reformulated as a problem of maximizing a monotone submodular function subject to a partition matroid constraint. Consequently, our problem MORE-R can be formulated as:

$$\max \sum_{n=1}^{N} U_n(Q_n(\chi))$$
$$s.t. \quad |\chi \cap E_i^l| \leq 1, \ \forall i = 1, 2, ..., m.$$

### F. Algorithm and Solution

We have proved that the objective function in MORE-R is monotone submodular. Based on the submodularity, we propose a simple but efficient greedy algorithm to handle MORE-R, which is illustrated in Algorithm 1.

From the Algorithm 1, we can see that, at each iteration, it computes all charging utility for all stop locations from all nodes in $O$. And we choose the grid which can acquire the maximum marginal value *i.e.*, the charging utility. Before the end of each iteration, we remove the nodes from $O$, that have met the required charging demands. Then the computation of useless marginal value from all nodes $O$ can be ignored. With this effective greedy scheme, we have the following theorem to give the guaranteed approximation ratio.

**Theorem 3.** *The near optimal Algorithm 1 for MORE-R problem can achieve approximation ratio of $\frac{1}{2}$. In addition, the time complexity of Algorithm 1 is $O(mn\Gamma)$, where $m$ is the number of time slots, $n$ is the number of nodes, $\Gamma$ is the number of grids obtained after grid partition.*

*Proof:* With the transformation and proof above, we know that our problem MORE-R is a problem of maximizing a monotone submodular function subject to a partition matroid

**Algorithm 1:** Near Optimal Algorithm for Transformed Problem MORE-R

**Input**: The nodes set $O = \{o_1, o_2, ..., o_n\}$, the deadline set $Y = \{\tau_1^l, \tau_2^l, ..., \tau_n^l\}$, the grids set $G = \{g_1, g_2, ..., g_\Gamma\}$, the time slots set $T_s = \{s_1, s_2, ..., s_m\}$, the charging utility model $U_i(x)$, the ground set $E = \bigcup_{h=1}^{m} E_h^l$ consists of candidate schedule $E_h^l$ and $z_h = 1, h = 1, 2, ..., m$.

**Output**: The schedule $\chi$ and effective charging quality.

1. $Q = 0; \chi = \varphi; O = O;$
2. $E_h = \varphi$, for $h = 1, 2, ..., m$;
3. **for** $h = 1, h \leq m, h++$ **do**
4.     **for** $k = 1, k \leq \Gamma, k++$ **do**
5.         $\Delta_Q(a_{hk}|\chi) = \sum_{n=1}^{N} X_h^k \times P_{kn} \times \Delta_t$;
6.         $a_{hk} = \arg\max_{a_{hk} \in E_{h^t}} \Delta_Q(a_{hk}|\chi)$;
7.         **if** $\Delta_Q(a_{hk}|\chi) == 0$ **then**
8.             break;
9.         $\chi = \chi \cup a_{hk}; E_h = E_h \cup a_{hk}$;
10.         **if** $||E_h|| == z_h$ **then**
11.             $E = E \setminus E_i^l$;
12.         **for** all $n$ $(1 \leq n \leq N)$ **do**
13.             **if** $U_n(\sum_{g_x \in \chi} X_x^n \cdot P_{xn} \cdot \Delta_t) == e_n$ **then**
14.                 remove the node $n$ from node set.

**Algorithm 2:** Skip-substitute Algorithm

**Input**: The stopped grids set $G^l = \{g_1^l, g_2^l, ..., g_m^l\}$ side length of grid $\delta$, the deadline set $D = \{\tau_1, \tau_2, ..., \tau_n\}$, the control parameter $\sigma$.

**Output**: The shortened traveling path $P^*$.

1. **for** all $l_k$ $(k = 1, 2, ..., m)$ **do**
2.     **if** all grids in $C(\overline{l_{k-1}l_k})$ are in $C(\overline{l_{k-1}l_{k+1}})$ **then**
3.         $P^* \leftarrow$ skip $l_k$ from $P$;
4.     **else**
5.         $l_s = l_k; l_t = l_{k+1}$;
6.         **while** $|l_s l_t| > \sigma$ **do**
7.             $l_k^l = (l_s + l_t)/2$;
8.             **if** all grids in $C(\overline{l_{k-1}l_k}) \cup C(\overline{l_k l_k^l})$ are in $C(\overline{l_{k-1}l_k^l})$ **then**
9.                 $l_s = l_k^l$;
10.             **else**
11.                 $l_t = l_k^l$;
12.         $P^* \leftarrow$ substitute $l_k$ by $l_k^l$ in $P$;

constraint. By the favourable property in [23], a typical greedy algorithm to a monotone submodular set function subject to a partition matroid constraint can acquire the approximation ratio of $\frac{1}{2}$. Hence, our algorithm follows. ∎

We claim that the multi-node and multiple charging schemes have been incorporated in Alg. 1. Due to the node-intensive deployment, the nodes can be charged when charger at the stop grids are within the range of the nodes. At each iteration, we traverse all the nodes for each time slot to realize the multi-node charging scheme. In addition, we introduce an example to show the multiple charging scheme. For example, if we get the charging scheduling $a_{hk}$ and $a_{(h+1)(k+1)}$ from ground set $E$, in which both the locations $g_k$ and $g_{k+1}$ are within the range of node $n$, then the node $n$ would be charged with two times. Therefore, the nodes can be charged repeatedly when charger are within the charging range even at different grids.

Note that, a randomised algorithm with optimal approximation ratio of $1 - \frac{1}{e}$ had been proposed in [24]. However, the randomised algorithm consists of pipage rounding technique and continuous greedy process, which generates high complexity at least for exponential type, especially the case of large nodes.

## V. SOLUTION TO MORE

In this section, we study the original problem MORE based on the analysis on MORE-R by taking the traveling time into consideration. In particular, we propose a *skip-substitute* algorithm on further optimizing the traveling path to save traveling time on the basis of solution $\chi$.

### A. Charging Route Constructing

Recall that after the area discretization, the charging power is approximated to be constant for nodes when the charger stops at any point in a grid. Therefore, as long as the traveling path of the charger intersects with a given grid, the charger can charge an ambient node with constant power at any point in the part of the traveling path lying in the grid. Let $P$ denote the traveling path, and $G^l = \{g_1^l, g_2^l, ..., g_m^l\}$ denote the charging positions ordered by time obtained by Algorithm 1. We aim at minimizing the traveling cost under the deadline constraints. With all above, the traveling path optimization problem can be formulated as

$$\min |P|$$
$$s.t. \ \forall g_k^l \in G^l; \ P \cap g_k^l \neq \varphi,$$

where $|P|$ is the tour length. The constraint indicates that all grids in $G^l$ are *path-covered* by $P$. Formally, let $C(P)$ denote the set of grids that are path-covered by $P$. As the grids in $G^l$ is ordered, our scheduling problem needs to find the shortest traveling path $P$ that visits the nodes in $G^l$ in the same order, i.e., $g_1^l, g_2^l, ..., g_m^l$.

### B. Skip-substitute Algorithm

In this subsection, we detail the skip-substitute algorithm, which basically consists of two steps. First, we traverse the grids $g_1^l, g_2^l, ..., g_m^l$ in $G^l$ in order and connect their center points. Second, we use a so-called skip-substitute method to shorten the tour.

Algorithm 2 shows the details of the algorithm. Generally, the basic idea of the algorithm is based on binary search. We iteratively search the skip path or substitute locations by binary search for all grids. Let $p_1, p_2, ..., p_m$ denote the stopped locations of the charger corresponding to $g_1^l, g_2^l, ..., g_m^l$ (i.e., $p_k$ is the center points of $g_k^l$ at the very begining, $k = 1, 2, ..., m$). For each location $p_k$ in $P$, we try to connect $p_{k-1}$ and $p_{k+1}$ and check whether $p_k$ is path-covered by segment $\overline{p_{k-1}p_{k+1}}$. If yes, we skip $p_k$; otherwise, we search another stopped location $p_k^l$ for the charger inside the segment $\overline{p_k p_{k+1}}$ using binary search with a granularity control parameter $\sigma$, under the constraint that $g_k$ should be path-covered by segment $\overline{p_{k-1}p_k^l}$. An example of Algorithm 2 is shown in Figure 5. First, the traveling path is initialized to be $\{S, p_1, p_2, p_3, p_4, S\}$. Then, the location $p_1$ is skipped because the grid $g_1$ is path-covered by $Sp_2$. The path is therefore updated to $\{S, p_2, p_3, p_4, S\}$. Next, we find that the grid $g_3$ is not path-covered by $p_2p_4$, and thus binary search is used to find a new path to substitute segment $\overline{p_2p_3}$. Specifically, path ①, ②, and ③ are tried and finally ③ is chosen as it path-covers $g_3$. We repeat the above skip-substitute process, and finally obtain the traveling path as shown in Figure 5.

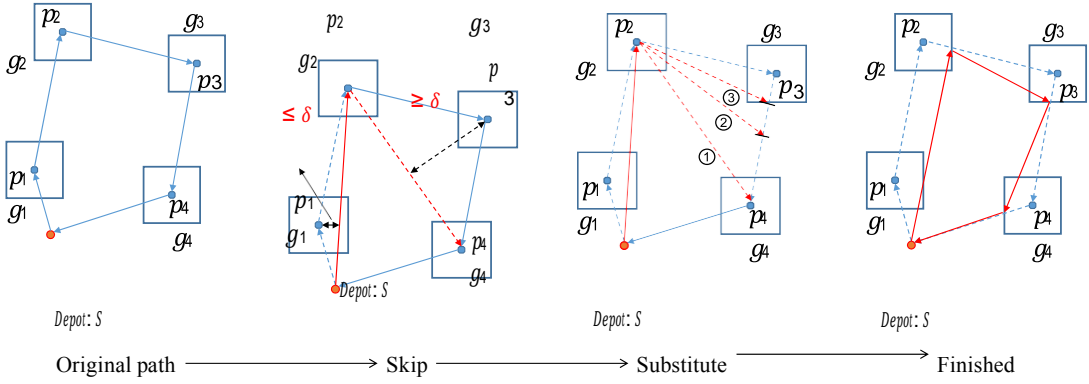

Fig. 5: Procedure of Algorithm 2.

By triangle inequality, we can easily derive the following lemma.

**Lemma 3.** *Each skip or substitute operation reduces the traveling length $|P|$.*

Let $P^*$ denote the optimal traveling path. With Lemma 3, we have

**Theorem 4.** $|P| \leq |P^*| + \sqrt{2}m\delta$. *In addition, the time complexity of Algorithm 2 is $O(m^2 \log \frac{\delta}{\sigma})$ where $m$ is the number of grids in $P$.*

*Proof:*
Given the set of grids $G_P = \{g_1, g_2, ..., g_m\}$, we use $P^* = \{S, p_1^*, ..., p_k^*, ..., p_m^*, S\}$ to denote the optimal traveling path, $P = \{S, p_1^l, ..., p_k^l, ..., p_m^l, S\}$ denote the traveling path obtained by Algorithm 2, where $p_0$ is the start and end point. The basic idea of the proof is to construct a detouring path in optimal traveling path $P^*$ to assist analysis. For any grids $g_k \in G_P$ ($p_k$ is the center point), we assume that the optimal stop location in $g_k$ is $p_k^*$, and $p_k^l$ is the stop location obtained by Algorithm 2. We construct the detour path by adding segment $\overrightarrow{p_k^* p_k^l p_k^*}$ into the optimal path $P^*$. An shown in Figure 6, in which $\{S, p_1^*, p_2^*, S\}$ is the optimal path, $\{S, p_1^l, p_2^l, S\}$ is the path obtained by *skip-substitute*, and $\{S, p_1^*, p_1^l, p_1^*, p_2^*, p_2^l, p_2^*, S\}$ is the detour traveling path. Similarly, the detour path is $\{S, p_1^* p_1^l p_1^*, ..., p_m^* p_m^l p_m^*, S\}$. By the triangle inequality, we have

$$|S, p_1^* p_1^l p_1^*, ..., p_m^* p_m^l p_m^*, S| \geq |P|.$$

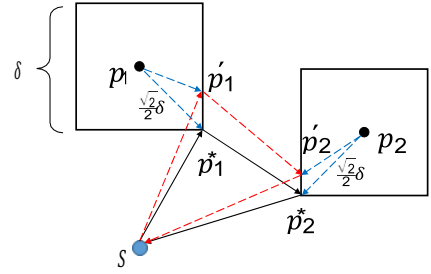

Fig. 6: Auxiliary illustration for proof of Theorem 4.

For the bounded grid, we have
$$|p_k^* p_k^l p_k^*| = 2|p_k^* p_k^l| \leq \sqrt{2}\delta.$$

Hence
$$|p_0, p_1^* p_1^l p_1^*, ..., p_m^* p_m^l p_m^*, p_0| \leq |P^*| + \sqrt{2}m\delta.$$

Thus, we obtain $|P| \leq |P^*| + \sqrt{2}m\delta$. ∎

Finally, we have to check whether or not the path $P = S, p_1^l, p_2^l, ..., p_l^l, S$, where $S$ is the depot for charger, obtained by Algorithm 2 is feasible given the constrained deadlines. Suppose the traveling time is predominant in the overall time consumption. Clearly, the traveling after the deadline budget $T$ is useless because of the deadline requirement. Then, the effective traveling is the path from the depot to a stop location until the deadline budget $T$. Let $P_x = \{S, p_1^l, p_2^l, ..., p_x^l, S\}$ denote the effective traveling path along the first $x$ stop locations in path $P$, ($p^l \in P$, $1 \leq x \leq l$), and the corresponding traveling time is $T_x = \sum_{i=0}^{x-1} |p_i p_{i+1}|$. Then, the effective traveling path $P^*$ under the constrained deadline $T$ is given by

$$P^* = \begin{cases} P_x, & if\ (T_x - T) \cdot (T_{x+1} - T) < 0; \\ P, & if\ T_p \leq T. \end{cases}$$

Consequently, we get the effective charging path $P^*$ under the constrained charging deadline.

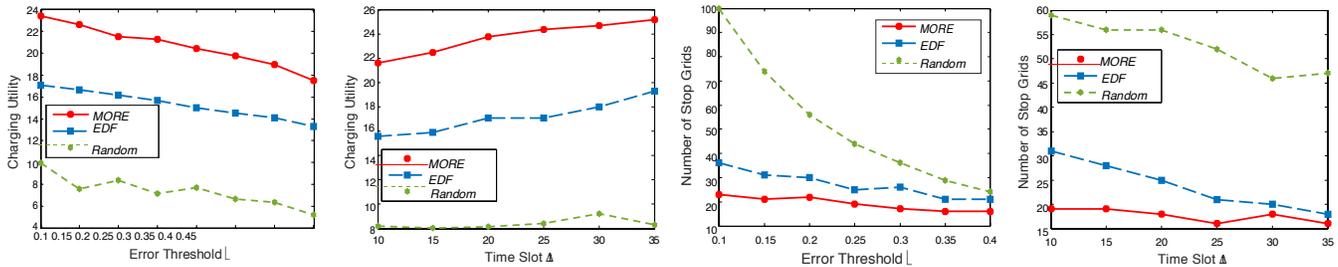

Fig. 7: Charging Utility vs. Error of grid partition. Fig. 8: Charging Utility vs. Time Slot. Fig. 9: Number of stopped grids vs. error $\lambda$. Fig. 10: Number of stopped grids vs. time slot.

## VI. EVALUATION

In this section, we conduct simulations to evaluate the performance of our algorithms. Particularly, we study the impacts of error $\lambda$ and slot $\Delta_t$ on charging utility, and the number of stop grids in terms of error $\lambda$ and slot $\Delta_t$. Without ambiguity, let MORE denote our proposed charging scheme in following context.

### A. Experimental Setup

Suppose there are 40 rechargeable nodes distributed on a plane of $50m \times 50m$, and a charger whose depot located at $[0, 0]$. Refer to [25], [26], we set $\alpha = 100$, $\beta = 10$, $D = 6m$ for the charging model, the error threshold $\lambda = 0.15$, the time slot $\Delta_t = 30s$, the energy of charging demand as random values in $[10J, 100J]$, and the charging deadline as random values in $[5min, 30min]$.

### B. Baseline Setup

As there is no related algorithm for scheduling chargers for deadline-driven multi-node charging, we suggest the *Early Deadline First (EDF)* scheduling algorithm which is applied in deadline-based data gathering [13], and *Random* algorithm for comparison. We modify *EDF* by changing the data gathering into wireless charging. At each iteration, the first one *Early Deadline First (EDF)* always chooses the nodes with the closest deadline, then it chooses the grids closest to the nodes until the deadline budget $T$. In this scheme, the charger performs charging only for one node at a time. In contrast, *Random* randomly selects a grid for each time slot to perform charging.

### C. Results and Performance

Firstly, we evaluate the effects of error threshold $\lambda$ and time slot $\Delta_t$ on charging utility. We set $\lambda$ and $\Delta_t$ be within $[0.1, 0.45]$ and $[10s, 35s]$, respectively. As shown in Figure 7 and 8, MORE outperforms *EDF* and *Random* for around 37.5% and 150% respectively.

In Figure 7, the charging utility for all schemes decreases as the error threshold $\lambda$ increases from 0.1 to 0.45. The charging utility for MORE drops around 27.1%, while around 21.2% for *EDF*. Furthermore, we can see that the gap between MORE and *EDF* decreases slightly. It indicates that MORE is more sensitive to $\lambda$ than *EDF*. The reason is, the increased error $\lambda$ results in a smaller number of grids, which would affect the selections in grids. However, *EDF* is not affected by it. Next, let us focus on Figure 8. In this figure, the charging utility for all schemes witnesses a growth as increasing time slot $\Delta_t$ from $10s$ to $35s$. MORE experiences the growth of 15.7% and *EDF* gets the growth of 23.1%. As $\Delta_t$ increases, the gap between our scheme and *EDF* decreases. The reason is that, the coarse-grained discretization of time slot would reduce marginal gain, *i.e.*, the charging utility in $\Delta_t$.

Next, we evaluate the impacts of time slot $\Delta_t$ and error $\lambda$ on the number of stopped grids for charger. We can observe in Figure 10 and 9 that MORE generates fewer stop grids for the charger, which generally leads to less traveling cost.

In Figure 9, the number of stop grids for all schemes decreases as increasing error $\lambda$ from 0.1 to 0.4. For *Random*, it goes through a significant fall in terms of stop grids. This can be explained by Lemma 1. The number of candidate grids would be reduced by increasing error $\lambda$. While for MORE, it witnesses a slight decrease. With the decreasing in number of grids, MORE always selects the optimal stop locations. For *EDF*, the stop locations won't change for it always first visits the node with the closest deadline. Furthermore, the similar reason could be used to explain the curves in Figure 10.

In summary, MORE can obtain the greatest charging utility with fewer stop locations for the charger compared with the other two comparison algorithms.

## VII. FIELD EXPERIMENTS

In this section, we conduct field experiments to validate the performance of MORE.

### A. Experiments Setup

As shown in Figure 11 and 12, we conduct our experiment on an indoor plane of $5m \times 5m$. Our testbed consists of 10 rechargeable sensor nodes and a TX91501 power transmitter produced by Powercast [27]. Note that we mount the charger on a robot car driven by Raspberry Pi to make it mobile. As the charging area of TX91501 power transmitters is roughly a sector with radius 4 and angle $120^o$, the robot car is controlled to rotate itself to mimic the omnidirectional charging. At the same time, an AP connecting to a laptop records the collected data from the nodes and reports it to the laptop. We set the charging distance threshold as $D = 1.2m$, the side length of grid partition $\delta = 0.3m$, and time slot length $\Delta_t = 10s$ based on our empirical results. The coordinates of ten rechargeable

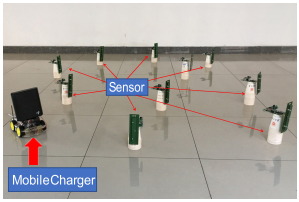 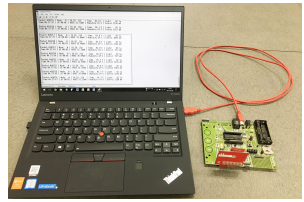 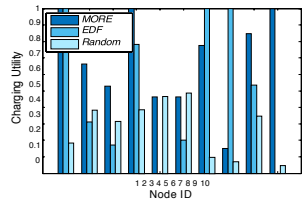 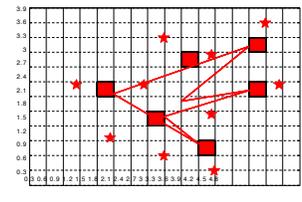

Fig. 11: The field placement.  Fig. 12: Data gathering module.  Fig. 13: Charging utility for each node.  Fig. 14: The traveling path for charger.

nodes are (1.15, 2.34), (1.75, 1.26), (2.35, 2.34), (3.54, 2.95), (3.55, 1.74), (4.74, 2.34), (4.48, 3.6), (2.7, 3.3), (2.7, 0.9), and (3.6, 0.6). The charging requests and its deadlines are randomly chosen from [1*J*, 5*J*] and [2*min*, 10*min*], respectively.

*B. Experiment Results*

Figure 14 shows the traveling path of the mobile charger for MORE. The depot is at (3, 2), the nodes are marked by red stars, and the red grids denote the obtained stop locations by MORE. Figure 13 shows the charging utility for each node for MORE, *EDF*, and *Random*. The overall charging utility for the 10 nodes of MORE, *EDF*, and *Random* are 6.89, 4.99, and 1.71, respectively. Our scheme outperforms *EDF*, *Random* by 37.9% and 146%, respectively. Note that the charging utility for node $7^{th}$ and $8^{th}$ of *EDF* is more than our scheme. The reason is that the nodes with early deadlines would always be charged first by *EDF*. However, the deadline-based *EDF* misses the $5^{th}$ and $10^{th}$ nodes and leads to their zero charging utility. In contrast, MORE can serve all the nodes and achieves higher overall charging utility.

## VIII. CONCLUSION

In this paper, we are the first to consider the problem of multi-node charging towards deadline-series named MORE. We first consider a relax version of MORE, *i.e.*, MORE-R, and by using the spatial and temporal discretization, we formulate MORE-R as maximizing a submodular function subject to a partition matroid constraint, which can be addressed by a greedy algorithm that achieves $\frac{1}{2}$ approximation ratio. Then, we present a skip-substitute algorithm to find the traveling path of the mobile charger, and thus address MORE. Finally, the simulations and field experiments show that MORE-R scheme outperforms the *EDF* scheme by 37.5% and 37.9%, respectively. In future work, we take the online case into consideration, which is more challenging and realistic.

ACKNOWLEDGMENTS

This research is partially supported by NSF of Jiangsu For Distinguished Young Scientist: BK20150030, NSFC with No. 61632010, 61232018, 61371118, China National Funds for Distinguished Young Scientists with No. 61625205, Key Research Program of Frontier Sciences, CAS, No. QYZDY-SSWJSC002, 61402009, 61672038, 61520106007 and NSF ECCS-1247944, NSF CMMI 1436786, and NSF CNS 1526638 and the National Science Foundation of USA under grants CNS-1526638.